# ANALYSIS OF THE NON-STATIONARY MODEL OF COUPLED OSCILLATORS WITH INDUCTIVE COUPLING

M.I. Ayzatsky[1], K.Yu. Kramarenko

National Science Centre "Kharkov Institute of Physics and Technology",
Kharkov, Ukraine
aizatsky@kipt.kharkov.ua

The model of coupled oscillators plays an important role in modern physics. It is used for description of various processes: from vibrations atoms in solid states to electromagnetic oscillations in slow-wave structures. The model with "short-range coupling" is the most widely used, for which a separate oscillator is coupled with two adjacent ones only. There are two main types of oscillators coupling: "capacitive" ("electric", "power") and "inductive" ("magnetic", "inertial"). In the first case, the coupling is proportional to the amplitudes of oscillations in the adjacent cells, in the second one – to the second derivative of these amplitudes. For numerical study of dynamics of a system that can be described by a model of coupled oscillators with an "inductive" coupling, it is necessary to find explicit expressions for the second derivatives of the amplitudes. To find these expressions, we propose to use the method that is based on the solution of difference equations. The results of the analysis of this method are given in the paper.

## *Introduction*

The model of coupled oscillators plays an important role in modern physics [1]. It is used for description of various processes: from vibrations of atoms in solid states [2,3,4,5], electromagnetic fields in RF devices, slow-wave structures and metamaterials [6,7,8,9,10,11,12,13,14,15,16,17,18,19] to biology and medicine [20,21,22]. The basis of the linear model is the system of coupled equations which in the frequency domain has a form

$$\left(\omega_{q_0}^{(n)2} + \alpha_{q_0,\omega}^{(n,n)} - \omega^2\right) e_{q_0,\omega}^{(n)} = \sum_{j=-\infty, j\neq n}^{\infty} e_{q_0,\omega}^{(j)} \alpha_{q_0,\omega}^{(n,j)}, \qquad (1)$$

where $e_{q_0,\omega}^{(n)}$ is the Fourier component of the complex oscillator amplitude $e_{q_0}^{(n)}(t)$, $\alpha_{q_0,\omega}^{(n,j)}$ are the coupling coefficients.

For simple type of coupling the coefficients $\alpha_{q_0,\omega}^{(n,j)}$ are the constants or are the simple functions of frequency $\omega$. For example, for study the linear vibrations of atoms in solid states the coupling coefficients can be considered as the constants, the same is true for the capacitive connected resonant electrical circuits. But in the case of inductive connected resonant electrical circuits they are proportional to the square of frequency $\omega^2$.

In many cases the coupled objects of the chain are the complex ones and each one can have several resonant frequencies. The most known network of this type is the chain of closed resonators that are connected through openings. The more complicated chains are the coupled resonator optical waveguides [16] and chains of closely spaced metal nanoparticles [17]. These chains are based on evanescent-field coupling[2] between the high-$Q$ modes of individual elements.

Each resonator of such chains has several (or infinite set) resonant frequencies. Despite the complexity of such coupled system it can be described by the simple system of coupled equations (1) . Indeed, if we consider, for example, the chain of closed microwave resonators, we can represent the electric field in the $n$-th resonator as

$$\vec{E}^{(n)} = \sum_q e_q^{(n)} \vec{E}_q^{(n)}(\vec{r}), \qquad (2)$$

---

[1] M.I. Aizatskyi, N.I.Aizatsky; aizatsky@kipt.kharkov.ua
[2] In a certain frequency domain

where $q = \{s, m, l\}$, $\vec{E}_q^{(n)}(\vec{r})$ is the field distribution for a $q$-mode. If we take the eigen oscillation with $q = q_0$ as the basic one, the coupling equations (1) for $e_{q_0,\omega}^{(n)}$ can be obtained [14,23,24,25].

Amplitudes of other modes ($q \neq q_0$) can be found by summing the relevant series

$$e_{q,\omega}^{(n)} = \frac{1}{\omega_q^{(n)2} - \omega^2} \sum_{j=-\infty}^{\infty} e_{q_0,\omega}^{(j)} \alpha_{q,\omega}^{(n,j)}. \tag{3}$$

For the chain of closed resonators in the frame of the Coupling Cavity Model (CCM) the coefficients $\alpha_{q,\omega}^{(n,j)}$ are electrodynamically strictly defined and can be calculated with necessary accuracy for arbitrary frequency [14,23-25]. The dependence of the coupling coefficients $\alpha_{q_0,\omega}^{(n,j)}$ on frequency is very complex (see Appendix). In the vicinity of the eigen frequency $\omega_{q_0}^{(n)}$ ($\omega \sim \omega_{q_0}^{(n)}$)[3] the coefficients $\alpha_{q_0,\omega}^{(n,j)}$ are continuous functions of frequency. But in the vicinity of the other eigen frequencies $\omega \sim \omega_q^{(n)}$, $q \neq q_0$ the coefficients $\alpha_{q,\omega}^{(n,j)}$, which can be calculated on the basis of CCM only with a finite number of coupling oscillators, have singularities[4] and the possibility of truncation of the sum in the system (1) and in equations for the coupling coefficients $\alpha_{q,\omega}^{(n,j)}$ becomes problematic [14,23-25]. This circumstance leads to the conclusion that system (1) is of practical importance only in the frequency domain in the vicinity of the eigen oscillation which was chosen as the basic one[5].

For the homogeneous chain of resonators ($\alpha_{q,\omega}^{(n,j)} = \alpha_{q,\omega}^{(n+j)}$) under the assumption that $e_{q_0,\omega}^{(n)} \sim \exp(in\varphi)$ from (1) we obtain the dispersive equation

$$\left(\omega_{q_0}^2 + \alpha_{q_0,\omega}^{(0)} - \omega^2\right) = \sum_{j=1}^{\infty} \alpha_{q_0,\omega}^{(j)} 2\cos(j\varphi), \tag{4}$$

that is true in the same frequency domain for which the system (1) is true.

The unknowns in the system of coupled equations (1) are the Fourier components of the complex oscillator amplitude $e_{q_0}^{(n)}$. Solving this system and making the inverse Fourier transform, we obtain the time dependent amplitudes $e_{q_0}^{(n)}(t)$.

Can we obtain on the basis of the equations (1) the system of differential equations for $e_{q_0}^{(n)}(t)$? Taking into account the complex dependence of the coefficients $\alpha_{q_0,\omega}^{(n,j)}$ on frequency, in general case the answer will be negative. But under some restrictions it is possible.

If we are interested in processes for which Fourier components are different from zero in the frequency range that include only the first passband of the system under consideration, there is a large frequency gap between the first and second passbands, we choose as the basic oscillations the ones that have the lowest frequencies, then for the conservative system of oscillators the coupling coefficients can be represented as

$$\alpha_{q_0,\omega}^{(n,j)} = \omega_{q_0}^{(n)2} \alpha_{n,j}^{(0)} + \omega^2 \alpha_{n,j}^{(1)} + \omega^4 \alpha_{n,j}^{(2)} / \omega_{q_0}^{(n)2} + \ldots \tag{5}$$

For system with coupling through absorption (passive or active) we have to take into account the odd powers of frequency in this expansion (see, for example, [26]). We shall not consider such systems.

If we can truncate this row as

$$\alpha_{q_0,\omega}^{(n,j)} \simeq \omega_{q_0}^{(n)2} \alpha_{n,j}^{(0)}, \tag{6}$$

---

[3] It is the frequency of the eigen oscillation which was chosen as the basic one.
[4] The reason of arising singularities explained in Appendix on the example of the simplest model.
[5] If the eigen oscillation which was chosen as the basic one has the lowest frequency, this domain spreads to zero frequency

then the coupling is called as "capacitive" ("electric", "power"). If the frequency dependence is playing an essential role in the coupling

$$\alpha_{q_0,\omega}^{(n,j)} \simeq \omega^2 \alpha_{n,j}^{(1)}, \qquad (7)$$

then the coupling is called as "magnetic" ("inertial").

Sometimes the first and the second terms are the same order and we have "mixed" coupling (see, for example, [17])

$$\alpha_{q_0,\omega}^{(n,j)} \simeq \omega_{q_0}^{(n)2} \alpha_{n,j}^{(0)} + \omega^2 \alpha_{n,j}^{(1)}. \qquad (8)$$

Restricting ourselves to this dependence, we can convert (1) into a system of differential equations for $e_{q_0}^{(n)}(t)$

$$\left(1-\alpha_{n,n}^{(1)}\right)\frac{d^2 e_{q_0}^{(n)}}{dt^2} + \omega_{q_0}^{(n)2}\left(1+\alpha_{n,n}^{(0)}\right)e_{q_0}^{(n)} = \omega_{q_0}^{(n)2} \sum_{j=-\infty, j\neq n}^{\infty} \alpha_{n,j}^{(0)} e_{q_0}^{(j)} - \sum_{j=-\infty, j\neq n}^{\infty} \alpha_{n,j}^{(1)} \frac{d^2 e_{q_0}^{(j)}}{dt^2}. \qquad (9)$$

The model with "short-range coupling" (nearest neighboring coupling), when each oscillator in the chain is coupled with two adjacent ones only, is used most widely

$$\left(1-\alpha_{n,n}^{(1)}\right)\frac{d^2 e_{q_0}^{(n)}}{dt^2} + \omega_{q_0}^{(n)2}\left(1+\alpha_{n,n}^{(0)}\right)e_{q_0}^{(n)} =$$
$$= \omega_{q_0}^{(n)2}\alpha_{n,n+1}^{(0)}e_{q_0}^{(n+1)} + \omega_{q_0}^{(n)2}\alpha_{n,n-1}^{(0)}e_{q_0}^{(n-1)} - \alpha_{n,n+1}^{(1)}\frac{d^2 e_{q_0}^{(n+1)}}{dt^2} - \alpha_{n,n-1}^{(1)}\frac{d^2 e_{q_0}^{(n-1)}}{dt^2} + f_n(t), \qquad (10)$$

where $f_n(t)$ is the external force acting on the $n$-th oscillator.

For finite number of oscillators this system can be analyzed by representing the solution as the sum of the eigen vectors with time dependence $\sim \exp(-i\Omega_i t)$ and with unknown amplitudes that have to be calculated (see, for example,[27,28,29]). This method is suitable for a small number of oscillators (resonators).

A second method is to approximate the differential equations (10) and determine the solution by incrementing step-by-step in time (see, for example, [30]).

A third method is to recast the equations (10) in the form of state equations

$$Y' = MY + F, \qquad (11)$$

where $Y$ is a vector with $2N$ state variable components, $N$ is the number of oscillators under consideration, $M$ is a $2N$ by $2N$ system matrix, and $F$ is a $2N$ vector of time-varying drive terms.

If we want to obtain a numerical solution of the system (10) the second and third methods are equivalent. The difference can be in the accuracy of difference schemes that are chosen for approximation the derivatives.

Application of the second and third methods for numerical analysis of the system (10) has a certain feature. Each equation of the system (10) has several highest derivatives[6]. So, we, for example, can not directly use the most common method of the transformation of the linear differential equation into a system of the first order equations by introducing a number of new variables (in our case $y_n = de_{q_0}^{(n)}/dt$). For both approaches this feature leads to the need for additional matrix transformations. This issue has not received sufficient coverage in the literature to date [10,11,27,28,29,30,31]. In its essence, these additional matrix transformations give explicit expressions for the second derivatives or the appropriative values of grid functions. For relatively small number of oscillators these matrix transformations can be easily performed. But there arise difficulties for the long chains of oscillators.

We propose to introduce in the procedure of analysis of the system (10) an intermediate transformation that clarify this issue and extend the range of the application of coupled oscillator

---

[6] If a signal spectrum is narrow these second derivatives of the amplitudes of neighbouring oscillators can be replaced by the amplitudes multiplied by the square of the resonant frequency

model. The results of applying of this transformation are given for two systems: an infinite chain of magnetically-coupled oscillators and a backward travelling wave (BTW) structure.

## 1. Infinite chain of oscillators

Let's consider an infinite chain of lossless "magnetically" coupled oscillators. The chain is described by the following system of the second-order differential equations

$$\frac{d^2 A_n}{dt^2} + \omega_0^2 A_n - \varepsilon\left(\frac{d^2 A_{n-1}}{dt^2} + \frac{d^2 A_{n+1}}{dt^2}\right) = F(t)\delta_{n,p}. \tag{12}$$

Here $A_n$ is the amplitude of the $n$-th oscillator, $\omega_0$ is the oscillator resonant frequency (all the oscillators are identical), $\varepsilon > 0$ is the coupling coefficient, $-\infty < n < \infty$, $F(t)$ is an external force that acts on the $p$-th oscillator.

For the case $F(t) \equiv 0$ and the time dependence of the amplitudes as $\exp(-i\omega t)$, the solution of the infinite system (12) can be written as

$$A_n = A_0 \rho^n, \tag{13}$$

where $\rho$ is the solution of a characteristic equation

$$\rho^2 - \rho\frac{\omega^2 - \omega_0^2}{\omega^2 \varepsilon} + 1 = 0. \tag{14}$$

By introducing

$$x_n = \frac{d^2 A_n}{dt^2}, \tag{15}$$

the system (12) can be rewritten as follows:

$$x_n - \varepsilon(x_{n-1} + x_{n+1}) = -\omega_0^2 A_n + F(t)\delta_{n,p}. \tag{16}$$

This is an inhomogeneous second-order difference equation with constant coefficients. The solution of this equation is

$$x_n = -\omega_0^2 \sum_{k=-\infty}^{\infty} X_{n-k} \cdot A_k + X_{n-p} F(t), \tag{17}$$

where $X_{n-k}$ is the fundamental solution

$$X_{n-k} = \begin{cases} \dfrac{g_2^{n-k}}{\sqrt{1-4\varepsilon^2}}, & n \geq k, \\ \dfrac{g_1^{n-k}}{\sqrt{1-4\varepsilon^2}}, & n \leq k, \end{cases} \tag{18}$$

$$g_{1,2} = \frac{1}{2\varepsilon} \pm \frac{1}{2\varepsilon}\sqrt{1-4\varepsilon^2}. \tag{19}$$

For small $\varepsilon$ ($\varepsilon \ll 1$) $g_2 \simeq \varepsilon$.

For $\varepsilon = \dfrac{1}{2} + \Delta$, $|\Delta| \ll 1$

$$\begin{aligned} |g_2| &\approx \begin{cases} 1 - 2\sqrt{|\Delta|}, & \Delta < 0, \\ \sqrt{1+4\Delta^2}, & \Delta > 0, \end{cases} \\ |g_1| &\approx \begin{cases} 1 + 2\sqrt{|\Delta|}, & \Delta < 0, \\ \sqrt{1+4\Delta^2}, & \Delta > 0. \end{cases} \end{aligned} \tag{20}$$

From this follows that the fundamental solution (18) is bounded at the infinity ($|g_2|<1$) for $\varepsilon<1/2$. We will consider this case.

Using the definition for $x_n$, we can write

$$\sqrt{1-4\varepsilon^2}\frac{d^2 A_n}{dt^2}+\omega_0^2 A_n+\omega_0^2\sum_{k=1}^{\infty}g_2^k\left(A_{n-k}+A_{n+k}\right)=g_2^{|n-p|}F(t). \tag{21}$$

The system of equations (12) and the system of equations (21) describe the same object: the infinite chain of identical magnetically coupled oscillators.

It can be shown that the homogeneous system of equations (21) has the solution of the form (13) with the same characteristic multiplier $\rho$.

It is useful to pay attention on several characteristic features of the obtained system (21).

Analysis of this system shows that instead of "magnetic" neighbouring coupling ("short connection") in the system (12) we obtained the "electrically" coupled oscillators with "long connection" (each oscillator is connected with all the others).

The external force that acts on the $p$-th oscillator in the chain of oscillators with "magnetic" neighbouring coupling transformed into the force that acts on all elements of the chain.

If we can truncate the sum in the system (21), it will be suitable to be directly transformed into the form which is easily solved numerically. As for small $\varepsilon$ ($\varepsilon<<1$) $g_2 \simeq \varepsilon$, then for $\varepsilon<1/2$ we can expect that the sum in the system (21) converges and can be truncated

$$\sum_{k=1}^{\infty}g_2^k\left(A_{n-k}+A_{n+k}\right)\simeq\sum_{k=1}^{K_c}g_2^k\left(A_{n-k}+A_{n+k}\right). \tag{22}$$

Bellow, on the example of the long chain of oscillators, we shall show that the sum in the system (21) really converges and the number of couplings $K_c$ that should be taken into account is determined by the value of $\varepsilon$.

Consider the chain of ($2N+1$) oscillators that can be described by the system of differential equations

$$\frac{dA_1}{dt}=Q_1;\ \left[1-\varepsilon g_2\right]\frac{dQ_1}{dt}+\omega_0^2 A_1+\omega_0^2\sum_{k=1}^{K_c}g_2^k A_{1+k}=g_2^N F(t),$$

$$\frac{dA_2}{dt}=Q_2;\ \left[1-2\varepsilon g_2\right]\frac{dQ_2}{dt}+\omega_0^2 A_2+\omega_0^2\sum_{k=1}^{K_c}g_1^k A_{2+k}+\omega_0^2 g_1 A_1=g_1^{N-1}F(t),$$

......

$$\frac{dA_{N+1}}{dt}=Q_{N+1};\ \left[1-2\varepsilon g_2\right]\frac{dQ_{N+1}}{dt}+\omega_0^2 A_{N+1}+\omega_0^2\sum_{k=1}^{K_c}g_1^k\left(A_{N+1-k}+A_{N+1+k}\right)=F(t), \tag{23}$$

......

$$\frac{dA_{2N}}{dt}=Q_{2N};\ \left[1-2\varepsilon g_2\right]\frac{dQ_{2N}}{dt}+\omega_0^2 A_{2N}+\omega_0^2\sum_{k=1}^{K_c}g_1^k A_{2N-k}+\omega_0^2 g_1 A_{2N+1}=g_1^{N-1}F(t),$$

$$\frac{dA_{2N+1}}{dt}=Q_{2N+1};\ \left[1-\varepsilon g_2\right]\frac{dQ_{2N+1}}{dt}+\omega_0^2 A_{2N+1}+\omega_0^2\sum_{k=1}^{K_c}g_1^k A_{2N+1-k}=g_1^N F(t).$$

If the external force start to act at $t=0$, this system will describe behavior of the infinitive chain of oscillators during a time interval, until the perturbations reach the chain end. If $N$ is large enough, we can explore the basic physical processes arising in the infinitive chain of oscillators. Bellow we consider the case $N=1000$.

As an input $F(t)$ we will use the signal of the type

$$F(t) = \exp(-i\omega t) \begin{cases} \sin\left(\dfrac{\pi}{2}\dfrac{t}{t_p}\right), & 0 \le t \le t_p, \\ 1, & t > t_p, \end{cases} \quad (24)$$

where the frequency $\omega = 2\pi/T$ of the input signal was supposed to have value

$$\frac{\omega^2}{\omega_0^2} = \frac{1}{1-2\varepsilon\cos\varphi}, \quad (25)$$

that gives such solutions of the characteristic equation (14)

$$\rho_{1,2} = \exp(\pm i\varphi). \quad (26)$$

For such excitation, the steady-state regime will be a uniform amplitude distribution with a phase shift $\varphi$ between the cells.

We used the Runge-Kutta method to find the numerical solution of the system (23). Instead time $t$ we used dimensionless value $\tau = \omega t$, so the quantity $\tau/2\pi$ represents the number of periods $T$ of the external force. In this section we assume $\varphi$ to be equal to $2\pi/3$. The calculation results for the long chain ($2N+1=2001$) with the small coupling ($\varepsilon = 0.01$) and different $K_c$ are shown in Fig. 1 and Fig. 2. We see that the amplitude distribution slightly changes with increasing $K_c$, while the phase distribution takes its expected form[7] only for $K_c \ge 2$. When the coupling became stronger, it is necessary to take into account the larger number of interacting oscillators (see Fig. 3-Fig. 6).

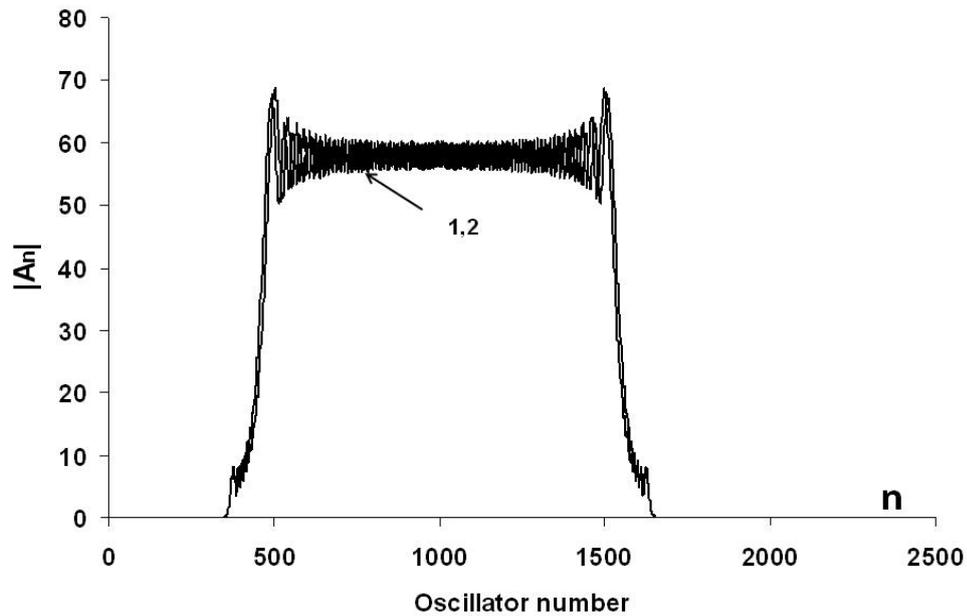

**Fig. 1 Amplitude distribution along the chain**
($1 - K_c = 1;\ 2 - K_c = 2;\ \varepsilon = 0.01,\ \tau = 2\pi\,10000$)

---

[7] Three parallel lines located at a distance φ from each other

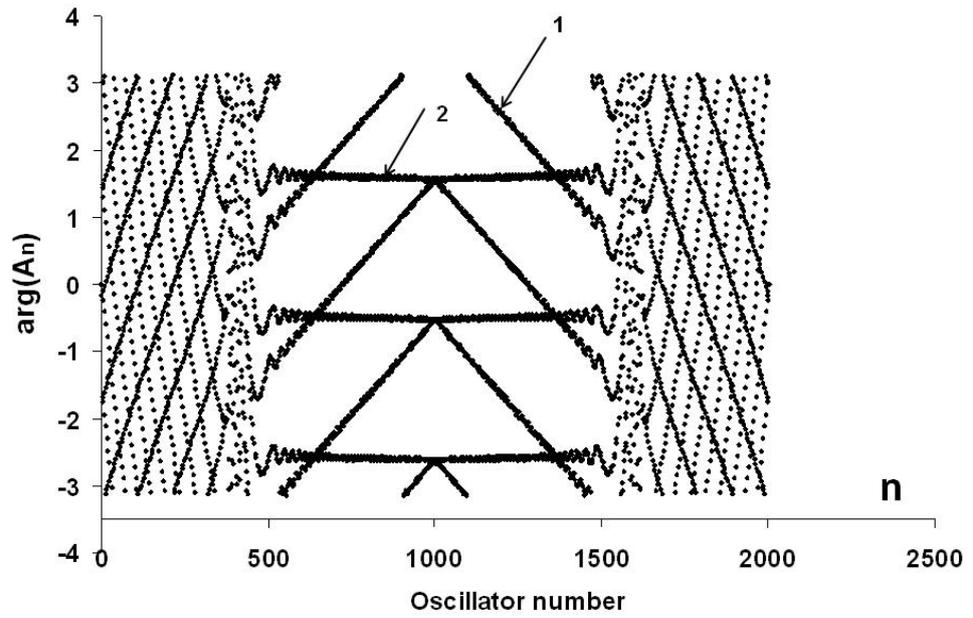

**Fig. 2 Phase distribution along the chain**
(1 - $K_c = 1$; 2 - $K_c = 2$; $\varepsilon = 0.01$, $\tau = 2\pi\,10000$)

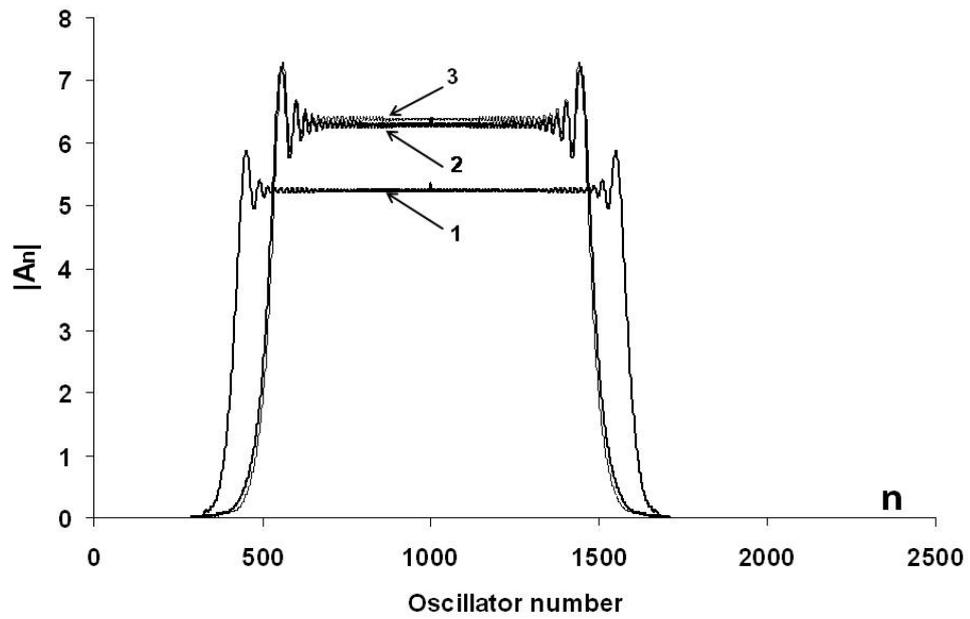

**Fig. 3 Amplitude distribution along the chain**
(1 - $K_c = 1$; 2 - $K_c = 2$; 3 - $K_c = 3$, $\varepsilon = 0.1$, $\tau = 2\pi\,1000$)

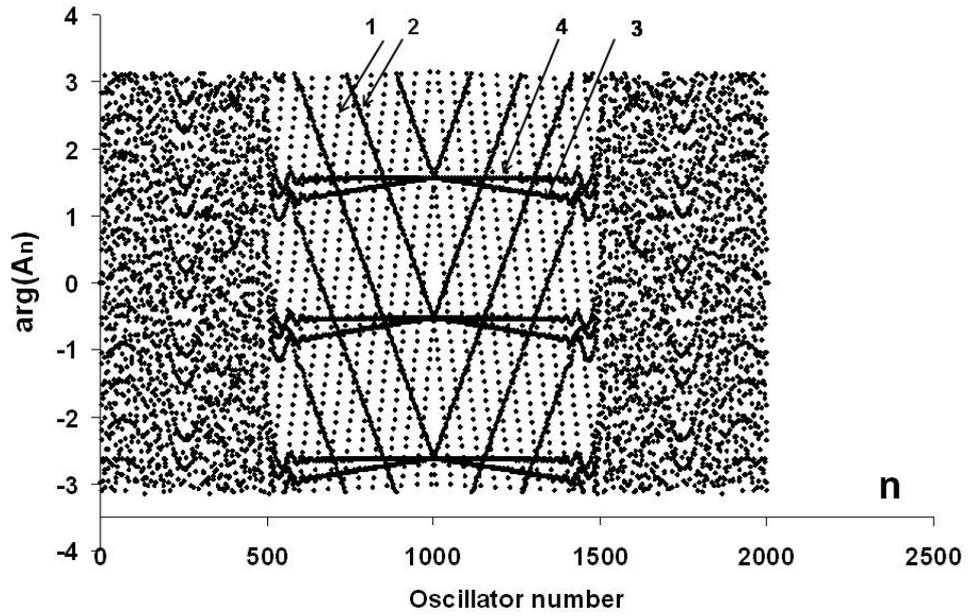

**Fig. 4 Phase distribution along the chain**
(1 - $K_c = 1$; 2 - $K_c = 2$; 3 - $K_c = 3$, ; **4 - $K_c = 4$;** $\varepsilon = 0.1$, $\tau = 2\pi\,1000$)

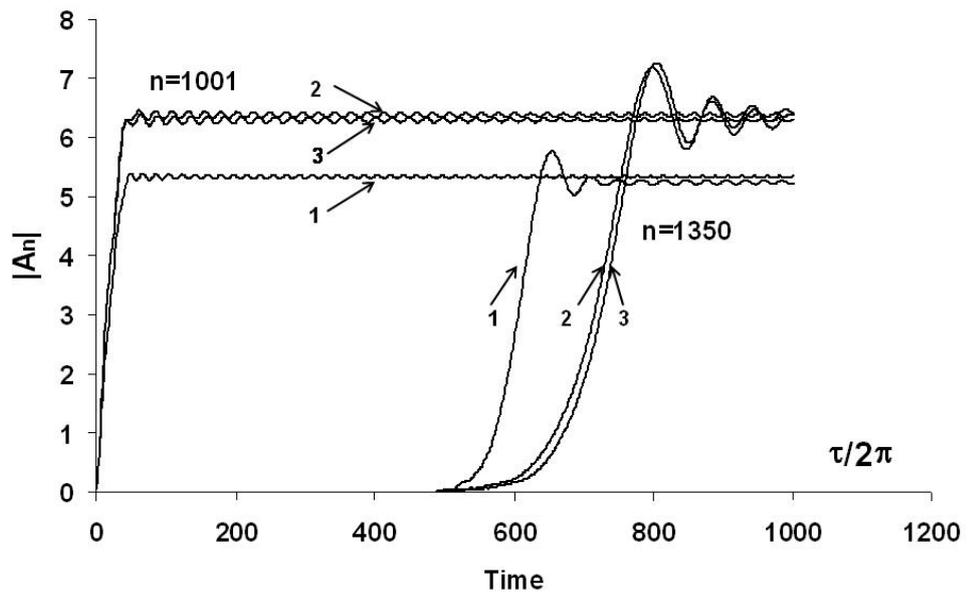

**Fig. 5 The time dependences of the amplitude of the oscillator with external force** ($n = 1001$) **and the amplitude of the oscillator at some distance from the source** ($n = 1350$)
(1 - $K_c = 1$; 2 - $K_c = 2$; 3 - $K_c = 3$, $\varepsilon = 0.1$)

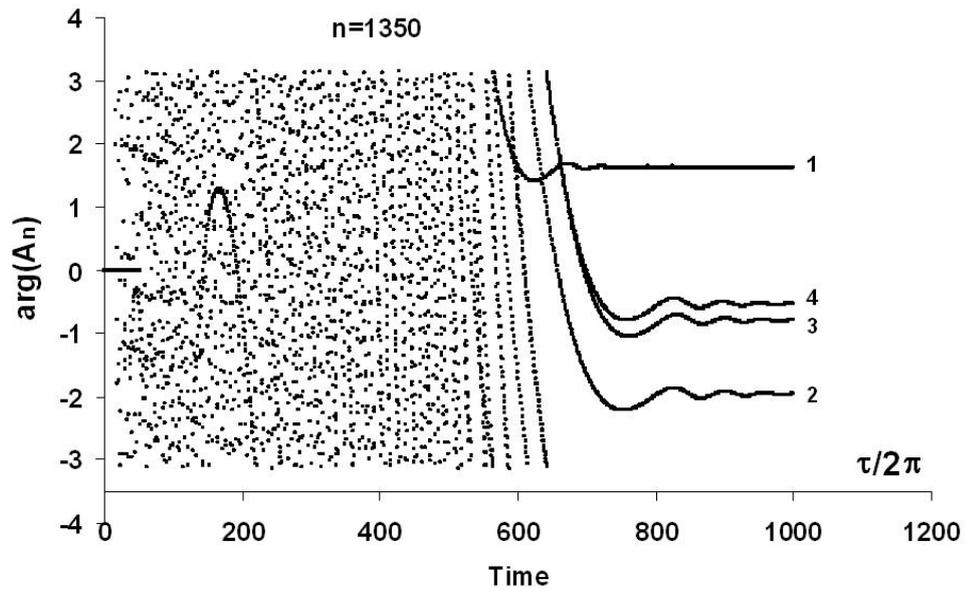

**Fig. 6 The time dependences of the phase of the oscillator at some distance from the source ($n = 1350$) (1 - $K_c = 1$; 2 - $K_c = 2$; 3 - $K_c = 3$; 4 - $K_c = 4$; $\varepsilon = 0.1$)**

## *2. Backward traveling wave section*

Let's consider the $N$ cells of disk-loaded waveguide (Fig. 7) The coupling between cells is magnetic, so the coupling slots (or holes) are located at the disks periphery.

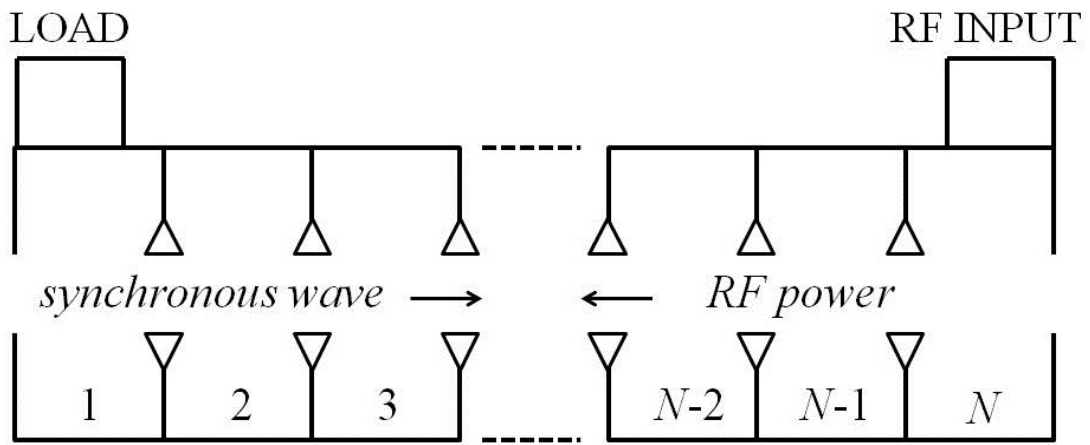

**Fig. 7 Backward traveling wave structure**

Considered structure is described by the following system of $N$ equations

$$\begin{cases} \dfrac{d^2 A_1}{d\tau^2}(1+\varepsilon) + A_1 \dfrac{\omega_1^2}{\omega_p^2} + \dfrac{\omega_1 \beta_1}{\omega_p Q_1} \dfrac{dA_1}{d\tau} - \varepsilon \dfrac{d^2 A_2}{d\tau^2} = 0, \\ \cdots \\ \dfrac{d^2 A_n}{d\tau^2}(1+2\varepsilon) + A_n \dfrac{\omega_0^2}{\omega_p^2} - \varepsilon \dfrac{d^2 A_{n-1}}{d\tau^2} - \varepsilon \dfrac{d^2 A_{n+1}}{d\tau^2} = 0, \\ \cdots \\ \dfrac{d^2 A_N}{d\tau^2}(1+\varepsilon) + A_N \dfrac{\omega_N^2}{\omega_p^2} - \varepsilon \dfrac{d^2 A_{N-1}}{d\tau^2} + \dfrac{\omega_N \beta_N}{\omega_p Q_N} \dfrac{dA_N}{d\tau} = -\dfrac{2\omega_N \beta_N}{\omega_p Q_N \theta} \dfrac{dU}{d\tau}, \end{cases} \quad (27)$$

where $U$ is the input RF pulse, $A_n$ is the amplitude of electric field in the $n$-th cell, $\tau = \omega_p t$, $\omega_p$ is the working frequency, $\omega_1$ and $\omega_N$ are the eigen frequencies of the couplers, $\omega_0$ is the eigen frequency of the cells, $\varepsilon$ is the coupling coefficient.

For the time dependence of the amplitudes and external signal as $\exp(-i\omega t)$ on the basis of the system (27) the parameters of resonators and its coupling were chosen to provide the phase shift between resonators $\varphi = 4\pi/5$ at the operating frequency $f_p = 2856$ MHz. The couplers parameters were chosen ($\beta_N/Q_N = \varepsilon\, \omega_p/\omega_N \sin\varphi$, $\beta_1/Q_1 = \varepsilon\, \omega_p/\omega_1 \sin\varphi$) to provide the absence of reflections from terminal cells at this frequency [32,33].

Denoting $d^2 A_n/d\tau^2 = \tilde{x}_n$ we can rewrite the set of equations (27) in the following form

$$\begin{cases} (1+\varepsilon)\tilde{x}_1 - \varepsilon \tilde{x}_2 = -A_1 \dfrac{\omega_1^2}{\omega_p^2} - \dfrac{\omega_1 \beta_1}{\omega_p Q_1} \dfrac{dA_1}{d\tau}, \\ \cdots \\ (1+2\varepsilon)\tilde{x}_n - \varepsilon(\tilde{x}_{n-1} + \tilde{x}_{n+1}) = -A_n \dfrac{\omega_0^2}{\omega_p^2}, \\ \cdots \\ (1+\varepsilon)\tilde{x}_N - \varepsilon \tilde{x}_{N-1} = -A_N \dfrac{\omega_N^2}{\omega_p^2} - \dfrac{\omega_N \beta_N}{\omega_p Q_N} \dfrac{dA_N}{d\tau} + \dfrac{2\omega_N \beta_N}{\omega_p Q_N \theta} \dfrac{dU}{d\tau}. \end{cases} \quad (28)$$

By analogy with the infinite structure, the solution of the system of equations (12) can be expressed through the fundamental solution

$$\dfrac{d^2 A_n}{d\tau^2} = -\tilde{X}_{n,1}\left(A_1 \dfrac{\omega_1^2}{\omega_p^2} + \dfrac{\omega_1 \beta_1}{\omega_p Q_1} \dfrac{dA_1}{d\tau}\right) - \dfrac{\omega_0^2}{\omega_p^2} \sum_{k=2}^{N-1} \tilde{X}_{n,k} \cdot A_k - \\ - \tilde{X}_{n,N}\left(A_N \dfrac{\omega_N^2}{\omega_p^2} + \dfrac{\omega_N \beta_N}{\omega_p Q_N} \dfrac{dA_N}{d\tau} - \dfrac{2\omega_N \beta_N}{\omega_p Q_N \theta} \dfrac{dU}{d\tau}\right). \quad (29)$$

Here the matrix $\tilde{X}$ is the fundamental solution of the system (28)

$$\begin{cases} (1+\varepsilon)\tilde{X}_{1,k} - \varepsilon \tilde{X}_{2,k} = \delta_{1,k}, \\ \cdots \\ (1+2\varepsilon)\tilde{X}_{n,k} - \varepsilon(\tilde{X}_{n-1,k} + \tilde{X}_{n+1,k}) = \delta_{n,k}, \\ \cdots \\ (1+\varepsilon)\tilde{X}_{N,k} - \varepsilon \tilde{X}_{N-1,k} = \delta_{N,k}, \end{cases} \quad (30)$$

where $1 \leq k \leq N$.

In general case all elements of the matrix $\tilde{X}$ are nonzero and the system (29) describes the interaction of individual resonators with all the other ones. Using in the system (29) truncated matrix $\bar{X}$ instead of the fundamental solution $\tilde{X}$, we can restrict the number of interacting resonators. For example, for the matrix

$$\bar{X} = \begin{pmatrix} \tilde{X}_{1,1} & \tilde{X}_{1,2} & \tilde{X}_{1,3} & 0 & 0 & 0 & 0 & \dots \\ \tilde{X}_{2,1} & \tilde{X}_{2,2} & \tilde{X}_{2,3} & \tilde{X}_{2,4} & 0 & 0 & 0 & \dots \\ \tilde{X}_{3,1} & \tilde{X}_{3,2} & \tilde{X}_{3,3} & \tilde{X}_{3,4} & \tilde{X}_{3,5} & 0 & 0 & \dots \\ 0 & \tilde{X}_{4,2} & \tilde{X}_{4,3} & \tilde{X}_{4,4} & \tilde{X}_{4,5} & \tilde{X}_{4,6} & 0 & \dots \\ 0 & 0 & \tilde{X}_{5,3} & \tilde{X}_{5,4} & \tilde{X}_{5,5} & \tilde{X}_{5,6} & \tilde{X}_{5,7} & \dots \\ \multicolumn{8}{c}{\dots\dots\dots\dots\dots\dots\dots\dots\dots\dots\dots} \end{pmatrix} \quad (31)$$

each resonator is coupled with the four neighboring ones[8]. For convenience, we shall mark the case when each resonator is coupled with the two neighboring ones as $K_c = 1$, when each resonator is coupled with the four neighboring ones as $K_c = 2$, when each resonator is coupled with the six neighboring ones as $K_c = 3$ and so on. The case of interacting of individual resonators with all the other ones we shall mark as $K_c = N$.

We used the Runge-Kutta method to find the numerical solution of the system (29). As an input we used the signal of the type (24).

The time dependence of the input and output amplitudes for the structure with $\varphi = 4\pi/5$ and $N = 100$ is shown in Fig. 8 ($\varepsilon = 0.02$, $\beta_g = v_g/c \approx -0.03$), Fig. 11 ($\varepsilon = 0.2$, $\beta_g = v_g/c \approx -0.17$). The time dependence of the amplitude of the reflected signal is shown in Fig. 9 and Fig. 12 for the same parameters. Deviations of phase distributions from a predetermined one ($\varphi_n = \varphi \times n$) for the same values of $\varepsilon$ are shown in Fig. 10 ($\tau = 2\pi \cdot 13000$) and Fig. 13 ($\tau = 2\pi \cdot 2200$).

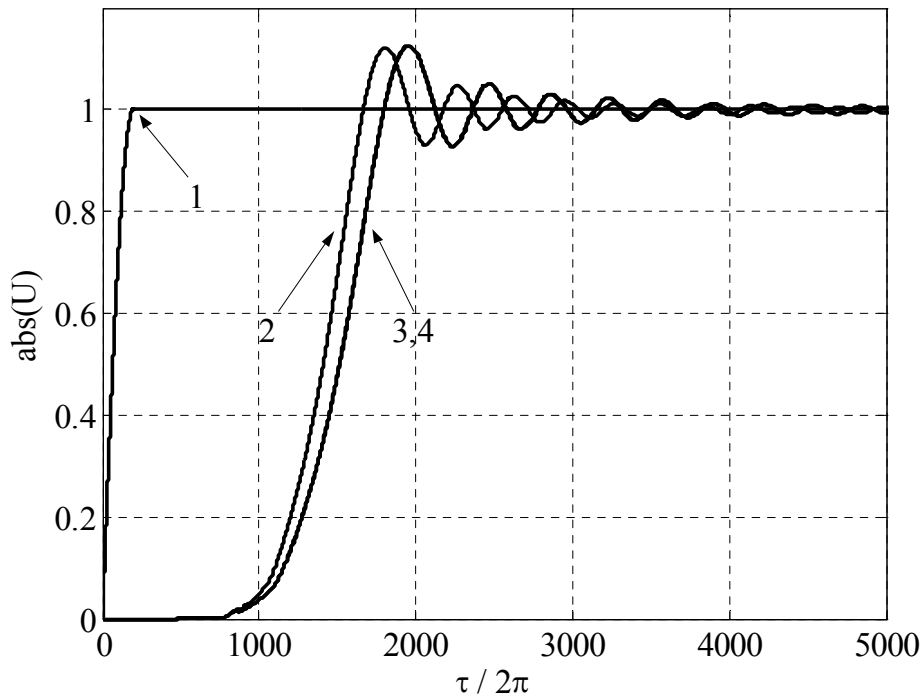

**Fig. 8 The time dependence of the input (1) and output amplitudes**
($2 - K_c = 1$; $3 - K_c = 2$; $4 - K_c = N$; $\varepsilon = 0.02$)

---

[8] Except for the terminal resonators

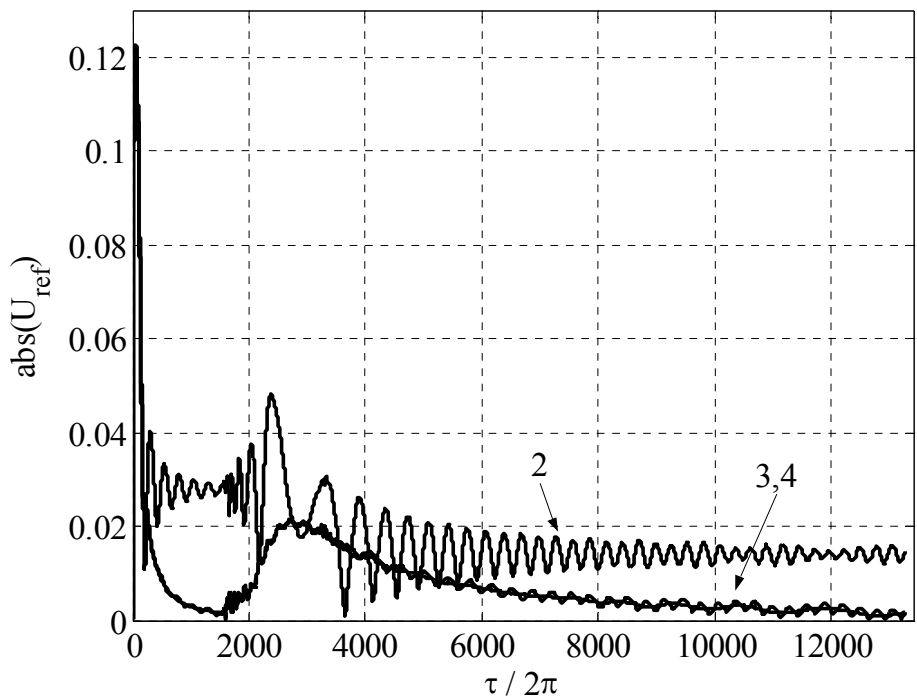

**Fig. 9 The time dependence of the reflected signal**
($2 - K_c = 1; 3 - K_c = 2; 4 - K_c = N$; $\varepsilon = 0.02$)

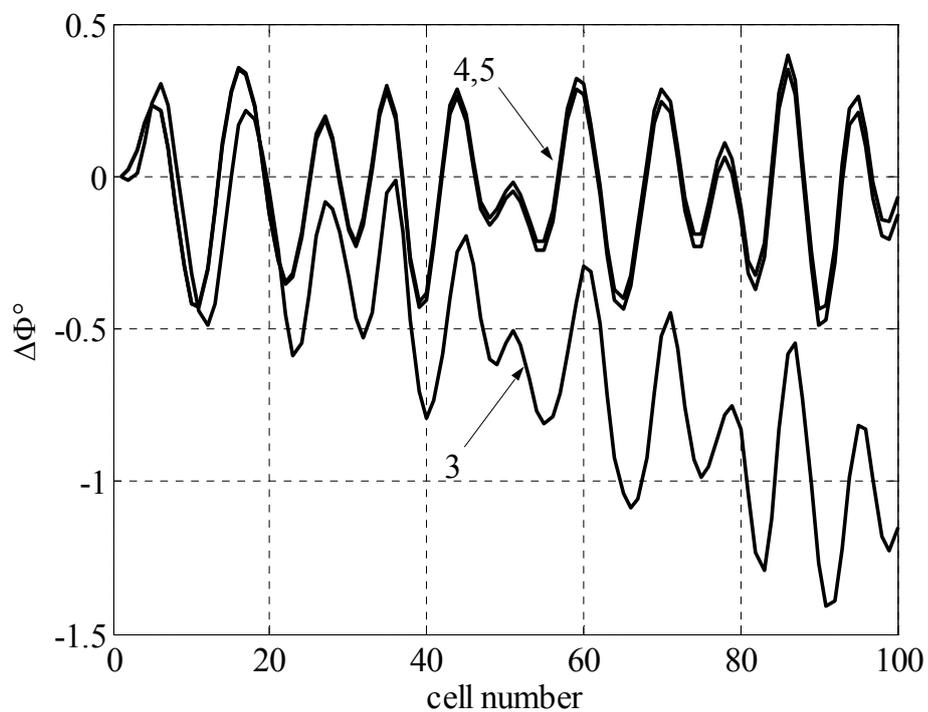

**Fig. 10 Deviation from a predetermined phase distribution**
($3 - K_c = 2; 4 - K_c = 3; 5 - K_c = N$, $\varepsilon = 0.02$, $\tau = 2\pi\,13000$)

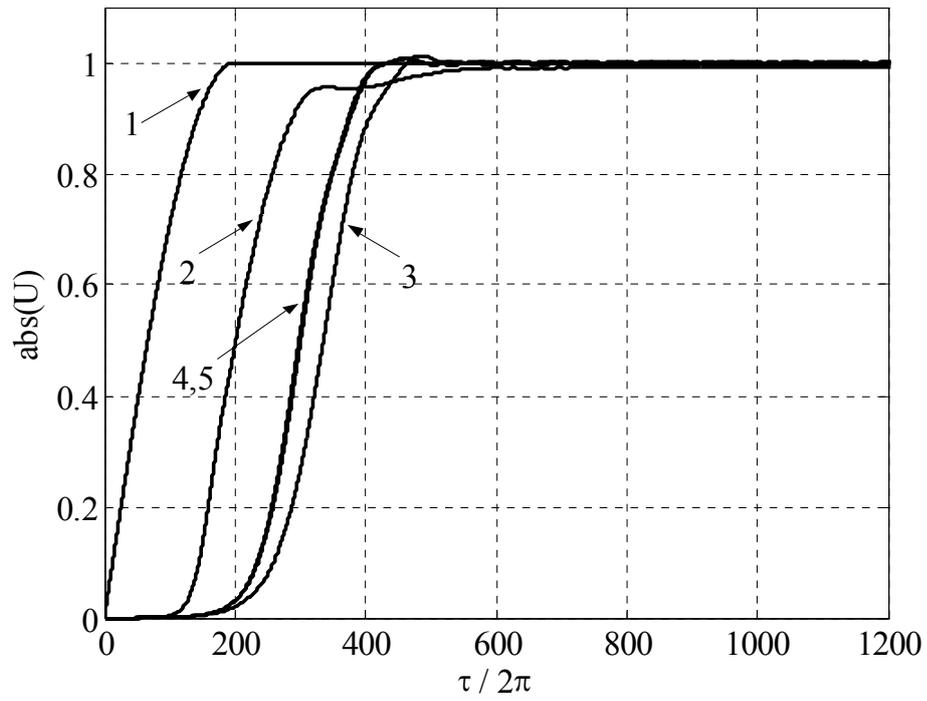

**Fig. 11 The time dependence of the input (1) and output amplitudes**
(2 - $K_c = 1$; 3 - $K_c = 2$; 4 - $K_c = 3$; 5 - $K_c = N$, $\varepsilon = 0.2$)

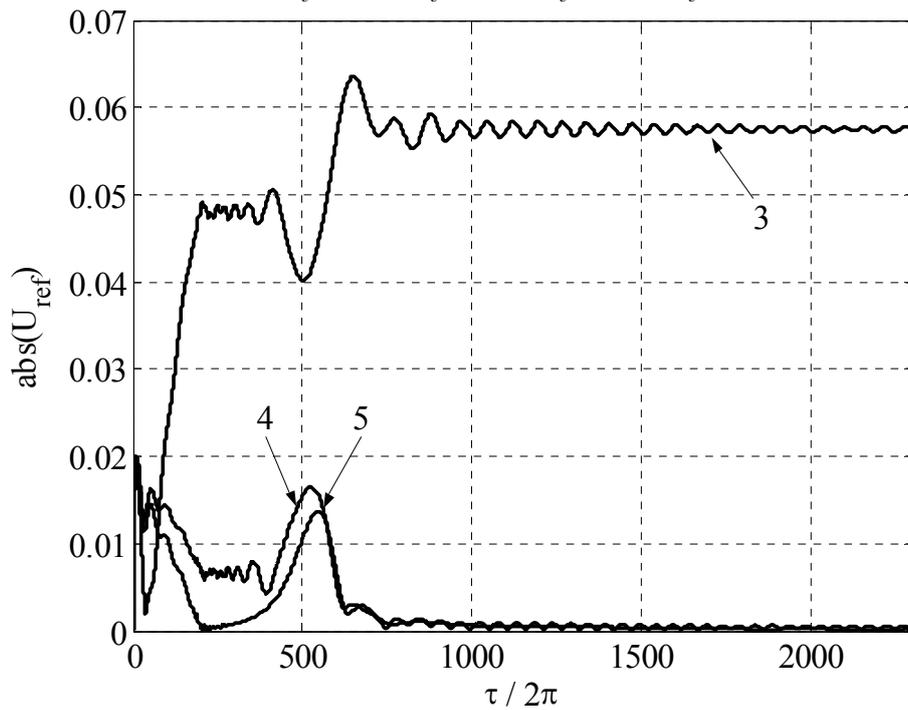

**Fig. 12 The time dependence of the reflected signal**
(3 - $K_c = 2$; 4 - $K_c = 3$; 5 - $K_c = N$; $\varepsilon = 0.2$)

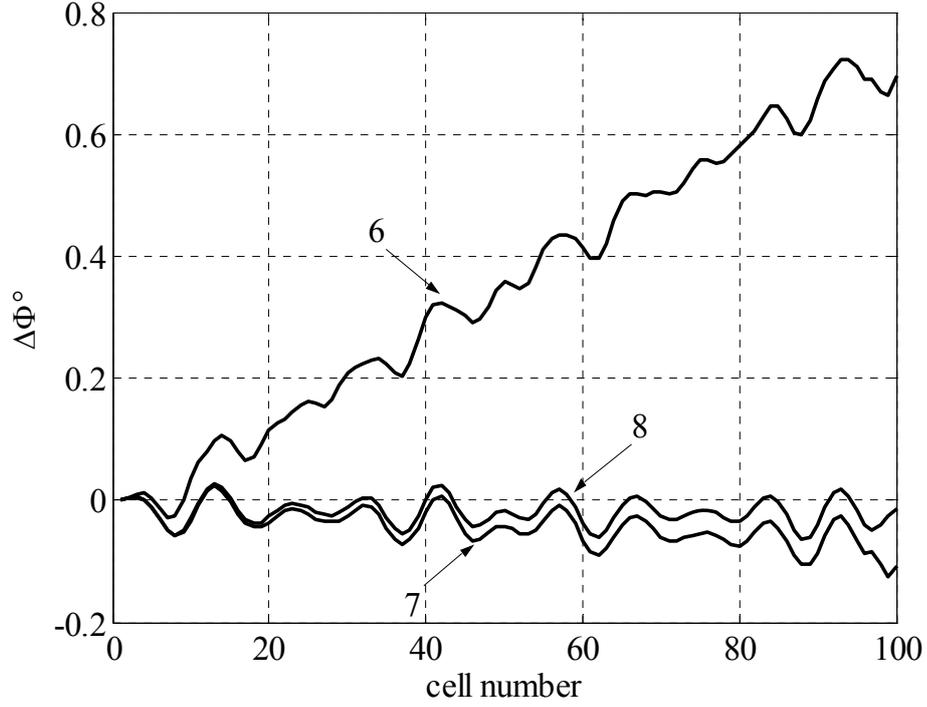

**Fig. 13 Deviation from a predetermined phase distribution**
(6 - $K_c = 5$; 7 - $K_c = 6$; 8 - $K_c = N$; $\varepsilon = 0.2$; $\tau = 2\pi\, 2200$)

Phase oscillations in Fig. 10 and Fig. 13 indicate that due to transient reflections from the couplers there is no completely steady state at the specified time.

Presented results show that the transients in the chain of magnetically coupled oscillators are sensitive to a mathematical model that used for numerical simulation. Especially strong influence of the value of coupling and the number of coupled resonators is observed for phase distributions (see Fig. 10, Fig. 13). Also the influence of coupling characteristics on the reflected signal can not be neglected. It is important for developing methods for tuning couplers and resonators [34,35] as they are based on the mathematical model of the coupled resonator chain [35].

## *Conclusion*

For numerical study of dynamics of a system that can be described by a model of coupled oscillators with an "inductive" coupling, we proposed to use the method that is based on the solution of difference equations. Based on this approach we analysed the influence of the value of coupling and the number of coupling oscillators on the characteristics of transients in the chain of magnetically coupled oscillators. It was shown that the transients in this chain are sensitive to a mathematical model that used for numerical simulation.

## *Appendix*

For the system consisting of two coupled resonators in the frame of the Coupling Cavity Model (CCM) the exact system of coupled equations can be obtained [36]

$$Z_A A_{1,\omega} = \omega_A^2 \Lambda_{11}(\omega) A_{1,\omega} + \omega_A^2 \Lambda_{12}(\omega) A_{2,\omega},$$
$$Z_A A_{2,\omega} = \omega_A^2 \Lambda_{22}(\omega) A_{2,\omega} + \omega_A^2 \Lambda_{21}(\omega) A_{1,\omega}, \qquad (32)$$

where $Z_A = (\omega^2 - \omega_A^2)$.

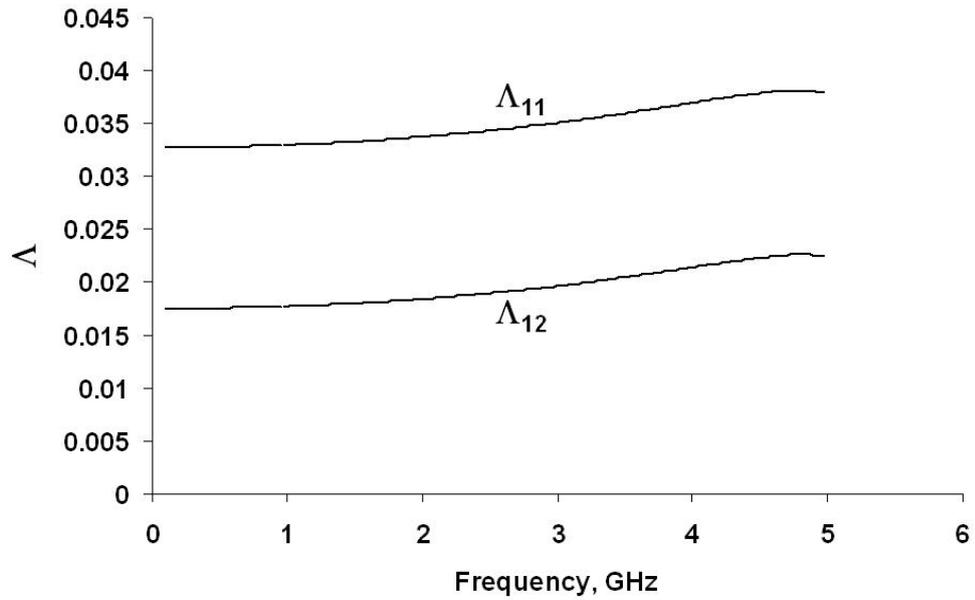

**Fig. 14** The frequency dependence of the coupling coefficients $\Lambda_{11}(\omega)$ and $\Lambda_{12}(\omega)$ for two identical resonators ($b = 4.3$ cm, $d = 3.0989$ cm) coupled through a cylindrical aperture of the radius $a = 1.5$ cm in the separating planar screen of the thickness $t = 0.4$ cm

The frequency dependences of the coupling coefficients $\Lambda_{11}(\omega)$ and $\Lambda_{12}(\omega)$ for different frequency ranges (near the $E_{010}$ and $E_{020}$ eigen frequencies), calculated in the frame of the CCM [36], are presented in Fig. 14 and Fig. 15. The eigen oscillations with $q = q_0 = (0,1,0)$ are chosen as the basic ones. The coupling coefficients $\Lambda_{11}(\omega)$ and $\Lambda_{12}(\omega)$ for frequency range near the $E_{020}$ modes have two singularities that ensure the existence of two eigen frequencies of coupled system in this frequency range.

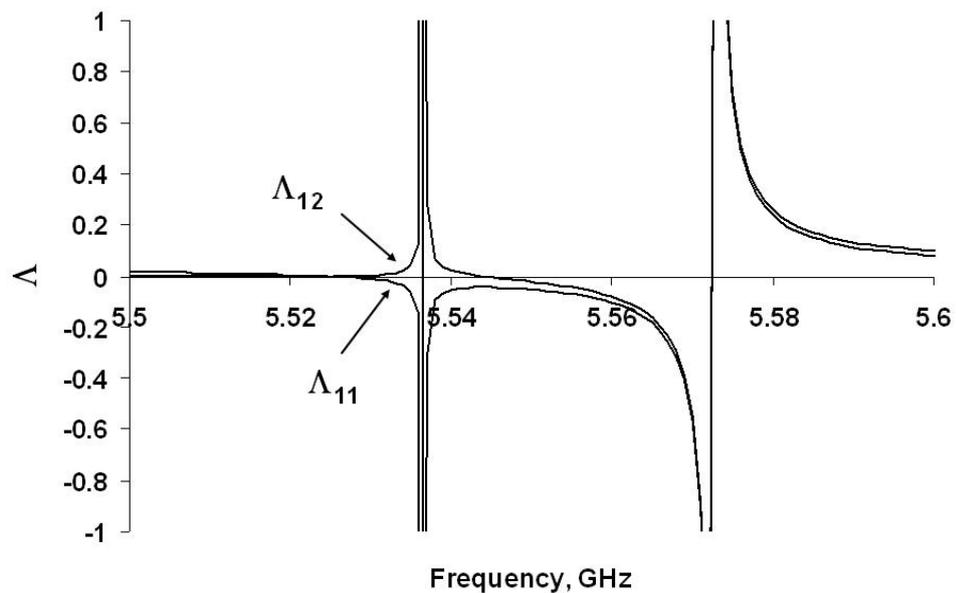

**Fig. 15** The frequency dependence of the coupling coefficients $\Lambda_{11}(\omega)$ and $\Lambda_{12}(\omega)$ for two identical resonators ($b = 4.3$ cm, $d = 3.0989$ cm) coupled through a cylindrical aperture of the radius $a = 1.5$ cm in the separating planar screen of the thickness $t = 0.4$ cm

To illustrate approach that gives us the coupling equations (32) and clarify the origin of singularities, we consider the model of two identical coupled resonators with choosing two eigen oscillations as the basic ones. The main equations of this model can be written as

$$Z_A A_{1,\omega} = \alpha_0 A_{1,\omega} + \alpha A_{2,\omega} + \bar{\alpha}_0 B_{1,\omega} + \bar{\alpha} B_{2,\omega},$$
$$Z_A A_{2,\omega} = \alpha_0 A_{2,\omega} + \alpha A_{1,\omega} + \bar{\alpha}_0 B_{2,\omega} + \bar{\alpha} B_{1,\omega},$$
(33)

$$Z_B B_{1,\omega} = \beta_0 B_{1,\omega} + \beta B_{2,\omega} + \bar{\beta}_0 A_{1,\omega} + \bar{\beta} A_{2,\omega},$$
$$Z_B B_{2,\omega} = \beta_0 B_{2,\omega} + \beta B_{1,\omega} + \bar{\beta}_0 A_{2,\omega} + \bar{\beta} A_{1,\omega},$$
(34)

where $Z_{A,B} = (\omega^2 - \omega_{A,B}^2)$.

From (34) we find

$$B_{1,\omega} D = (\bar{\beta}_0 A_{1,\omega} + \bar{\beta} A_{2,\omega})(Z_B - \beta_0) + \beta(\bar{\beta}_0 A_{2,\omega} + \bar{\beta} A_{1,\omega}),$$
$$B_{2,\omega} D = (\bar{\beta}_0 A_{2,\omega} + \bar{\beta} A_{1,\omega})(Z_B - \beta_0) + \beta(\bar{\beta}_0 A_{1,\omega} + \bar{\beta} A_{2,\omega}),$$
(35)

where $D = (Z_B - \beta_0)^2 - \beta^2$.

Substituting (35) into (33), we get

$$Z_A A_{1,\omega} = \omega_A^2 \Lambda_{11} A_{1,\omega} + \omega_A^2 \Lambda_{12} A_{2,\omega},$$
$$Z_A A_{2,\omega} = \omega_A^2 \Lambda_{22} A_{2,\omega} + \omega_A^2 \Lambda_{21} A_{1,\omega},$$
(36)

where

$$\Lambda_{11} = \Lambda_{22} = \omega_A^{-2}\left[\alpha_0 + \frac{\bar{\alpha}_0}{D}\{\bar{\beta}_0(Z_B - \beta_0) + \beta\bar{\beta}\} + \frac{\bar{\alpha}}{D}\{\bar{\beta}(Z_B - \beta_0) + \beta\bar{\beta}_0\}\right],$$
$$\Lambda_{12} = \Lambda_{21} = \omega_A^{-2}\left[\alpha + \frac{\bar{\alpha}_0}{D}\{\bar{\beta}(Z_B - \beta_0) + \beta\bar{\beta}_0\} + \frac{\bar{\alpha}}{D}\{\bar{\beta}_0(Z_B - \beta_0) + \beta\bar{\beta}\}\right].$$
(37)

If $|\bar{\alpha}_0/D| \ll 1$, $|\bar{\alpha}/D| \ll 1$ [9] then, neglecting the quadratic terms of smallness, we get the system of two coupled oscillators

$$Z_A A_{1,\omega} = \alpha_0 A_{1,\omega} + \alpha A_{2,\omega},$$
$$Z_A A_{2,\omega} = \alpha_0 A_{2,\omega} + \alpha A_{1,\omega},$$
(38)

and the eigen frequencies of the coupled system are determined by the equation

$$Z_A - \alpha_0 = \pm\alpha.$$
(39)

In this case the influence of the second resonances is negligible.

At the frequencies when $|D| \sim 0$ the coupling coefficients $\Lambda$ have singularities. But these singularities do not lead to unphysical results. Indeed, if $|\bar{\alpha}_0/D| \gg 1$, $|\bar{\alpha}/D| \gg 1$ we can neglect in the system (36) the terms that proportional $D$ and we get

$$\left[\bar{\alpha}_0\{\bar{\beta}_0(Z_B - \beta_0) + \beta\bar{\beta}\} + \bar{\alpha}\{\bar{\beta}(Z_B - \beta_0) + \beta\bar{\beta}_0\}\right]A_{1,\omega} = $$
$$-\left[\bar{\alpha}_0\{\bar{\beta}(Z_B - \beta_0) + \beta\bar{\beta}_0\} + \bar{\alpha}\{\bar{\beta}_0(Z_B - \beta_0) + \beta\bar{\beta}\}\right]A_{2,\omega},$$
$$\left[\bar{\alpha}_0\{\bar{\beta}_0(Z_B - \beta_0) + \beta\bar{\beta}\} + \bar{\alpha}\{\bar{\beta}(Z_B - \beta_0) + \beta\bar{\beta}_0\}\right]A_{2,\omega} = $$
$$-\left[\bar{\alpha}_0\{\bar{\beta}(Z_B - \beta_0) + \beta\bar{\beta}_0\} + \bar{\alpha}\{\bar{\beta}_0(Z_B - \beta_0) + \beta\bar{\beta}\}\right]A_{1,\omega}.$$
(40)

This system has the non-trivial solutions if its determinant equals to zero

$$\left[\bar{\alpha}_0\{\bar{\beta}_0(Z_B - \beta_0) + \beta\bar{\beta}\} + \bar{\alpha}\{\bar{\beta}(Z_B - \beta_0) + \beta\bar{\beta}_0\}\right]^2 = $$
$$= \left[\bar{\alpha}_0\{\bar{\beta}(Z_B - \beta_0) + \beta\bar{\beta}_0\} + \bar{\alpha}\{\bar{\beta}_0(Z_B - \beta_0) + \beta\bar{\beta}\}\right]^2.$$
(41)

---

[9] It is takes place if the frequency $\omega$ is not close to the eigen frequency $\omega_B$

This can be rewritten as

$$(Z_B - \beta_0) = \pm\beta. \qquad (42)$$

Such characteristic equation can be obtained from (34) by neglecting the terms concerning the "A" mode.

So, the presence of singularities in the coupling coefficients $\Lambda$ gives possibility to correctly describe the oscillatory features of the complex systems that have many resonant frequencies with using the simpler system of equations.